\documentclass[]{spie}  

\usepackage{amsmath,amsfonts,amssymb}
\usepackage{graphicx}
\usepackage{placeins}
\usepackage{Euclid}
\usepackage[colorlinks=true, allcolors=blue]{hyperref}

\title{Euclid Near Infrared Spectro-Photometer: spatial considerations on H2RG detectors interpixel capacitance and IPC corrected conversion gain from on-ground characterization}

\author[a]{Le Graët J.}
\author[a]{Secroun A.}
\author[b]{Barbier R.}
\author[a]{Gillard W.}
\author[a]{Clémens J-C.}
\author[b]{Conseil S.}
\author[a]{Escoffier S.}
\author[b]{Ferriol S.}
\author[b]{Fourmanoit N.}
\author[c]{Kajfasz E.}
\author[a]{Kermiche S.}
\author[b]{Kubik B.}
\author[b]{Smadja G.}
\author[a]{Zoubian J.}
\author[$_{\,}$]{on behalf of the Euclid Consortium}

\affil[a]{Aix Marseille Univ, CNRS/IN2P3, CPPM, Marseille, France}
\affil[b]{IP2I, CNRS, Lyon University, 4 rue E. Fermi, 69100 Villeurbanne, France}
\affil[c]{Aix Marseille Univ, CNRS/IN2P3, CPPM, IPhU, Marseille, France}

\authorinfo{Further author information: Send correspondence to Jean Le Graët\\E-mail: legraet@cppm.in2p3.fr, Telephone: +33 (0)4 91 82 72 90\\ }

\begin{document} 
\maketitle

\begin{abstract}






\Euclid is a major ESA mission scheduled for launch in 2023-2024 to map the geometry of the dark Universe using two primary probes, weak gravitational lensing and galaxy clustering. \Euclid's instruments, a visible imager (VIS) and an infrared spectrometer and photometer (NISP) have both been designed and built by Euclid Consortium teams. The NISP instrument will hold a large focal plane array of 16 near-infrared H2RG detectors, which are key elements to the performance of the NISP, and therefore to the science return of the mission.

\Euclid NISP H2RG flight detectors have been individually and thoroughly characterized at Centre de Physique des Particules de Marseille (CPPM) during a whole year with a view to producing a reference database of performance pixel maps. Analyses have been ongoing and have shown the relevance of taking into account spatial variations in deriving performance parameters. This paper will concentrate on interpixel capacitance (IPC) and conversion gain. First, per pixel IPC coefficient maps will be derived thanks to single pixel reset (SPR) measurements and a new IPC correction method will be defined and validated. Then, the paper will look into correlation effects of IPC and their impact on the derivation of per super-pixel IPC-free conversion gain maps. Eventually, several conversion gain values will be defined over clearly distinguishable regions.

\end{abstract}


\section{INTRODUCTION}
\label{sec:intro}  

\Euclid is a medium class ESA mission\cite{Racca_2016} due to launch in 2023-2024. The \Euclid mission is primarily dedicated to cosmology and aims at unveiling the nature of dark energy and dark matter. \Euclid will observe $15 \, 000$\,deg$^2$ of extragalactic sky using weak gravitational lensing and galaxy clustering – including baryonic acoustic oscillations and redshift space distortions – as cosmological probes. \Euclid’s instruments, a visible imager (VIS) and a near-infrared spectro-photometer (NISP), have both been designed and built by the Euclid Consortium teams. 

The NISP instrument\cite{Maciaszek2016}, operating in the near-IR spectral band ([0.9, 2.0]\,\micron,) is dedicated to the measurement of about 40 millions galaxies' redshifts, gaining a factor 10 compared to previous similar missions, providing an unprecedented 3D distribution of galaxies in the Universe. NISP's large focal plane array holds 16 infrared detectors with a 2.3\,\micron \, cutoff. Since NISP will mostly observe very low signals from galaxies (about 2\,photon\,s$^{-1}$), detectors with high quantum efficiency (QE) and ultra-low readout noise and dark current are keys to the overall performance of the instrument, and therefore to the science return of the mission.
Many past and current ground and space missions have integrated HxRG detectors for their high performance : for instance H2RG onboard JWST\cite{JWST}, H4RG onboard the future \textit{Roman} space telescope\cite{WFIRST} mission, or the MOONS\cite{moons} instrument of the Very Large Telescope (VLT) and H1RG onboard ARIEL\cite{Ariel}, an ESA mission devoted to exoplanets atmospheric characterization. H2RG detectors from Teledyne have also been chosen for the \Euclid mission.

NISP detectors consist of a matrix of $2040 \times 2040$ science pixels surrounded by a 4 pixel wide ring of reference pixels used to minimize correlated noise~\cite{kubik_2014}. The detectors are read out by sidecar cold electronics provided by Teledyne on 32 channels at 100\,kHz. Due to the nondestructive readout, each acquisition provides a ramp, i.e. successive frames taken every 1.45 seconds, each of which made up of ADU (Analog Digital Unit) values for all the pixels of the matrix. The 20 $2048 \times 2048$ pixel detectors (16 flight detectors + 4 spares) have been selected by NASA among 60 detectors on the basis of a figure of merit combining QE, read noise and dark current. In addition to this initial selection phase, the on-ground characterization of the selected flight detectors that has taken place at CPPM, in Marseilles France, aimed to provide performance maps at the pixel level of parameters such as linearity or conversion gain\cite{secroun_2016}. 

While detectors have been pushed to increasingly higher performance, various flaws came to light, such as the nonlinearity of the pixel response\cite{plazas_2017}, interpixel capacitance (IPC)\cite{finger_conversion_2005-1} and persistence\cite{Smith_2008}. These shortcomings may directly impact the calibration of the detectors response. For instance, it has been shown\cite{moore_quantum_2006} that IPC induces a systematic error on the conversion gain measured by the standard mean variance method. Error on the conversion gain will in turn impact the calculation of readout noise, dark current, or QE. These errors result directly from correlations existing between the pixels of a detector and consequently from methods used to derive the parameters. They need to be properly taken into account when deriving performance maps.

For this purpose, \Euclid NISP H2RG flight detectors have been individually and thoroughly characterized at CPPM during a whole year, varying parameters such as temperature, flux level, bias, illumination history, etc. to cover varied flight configurations\cite{barbier_2018} at the price of producing $500$\,TB of raw data.
From initial analyses, several parameters such as IPC, persistence and conversion gain have shown large spatial variations and interdependence. If using mean values of these parameters is usually the rule, looking into spatial variations could provide missions with an insight into their impact on science data. Depending on the precision needed, missions will be able to choose from a mean value over the detector down to a per pixel value of the different parameters.
This paper is divided into two parts: the first part concentrates on IPC and how to properly apply per pixel IPC correction to measurement data, whereas the second part describes a method to derive IPC corrected conversion gain on superpixels.

\section{INTERPIXEL CAPACITANCE}
\label{sec:ipc}

IPC describes electrical crosstalk between adjacent pixels. Precisely, the proximity of pixels in HxRGs produces an electrostatic coupling that causes the signal detected by a pixel to spread to its nearest neighbors. This effect results mostly from an electrical field created in the intermediate layer separating the photosensitive layer and the multiplexer. It does not affect the number of charges in the pixel but the voltage at its extremity making it non-stochastic and so fully characterizable.

As a major consequence, the presence of IPC is known to lead to a deterioration of the point spread function (PSF): recent studies \cite{Kannawadi_2016} estimate this effect to 5\,\% of the PSF size. The potential impact on the \Euclid PSF has been taken into consideration but it is expected that it will be negligible since NISP's PSF is under-sampled. As another consequence, the standard variance vs. signal method is biased by the coupling and underestimates the conversion gain.  Ways of taking IPC into account in the calculation of \Euclid detectors' conversion gain will be presented in Sect.~\ref{sec:conversiongain}.

IPC is usually defined by a kernel~\cite{moore_quantum_2006,finger_conversion_2005-1} filled with the percentage of signal detected on a pixel that originates from the central pixel of the kernel. The sum of the contributions of all the pixels of the kernel should therefore be 100\% because IPC is flux conserving. The $\alpha$ coefficients commonly used in other papers\cite{moore_quantum_2006} are just those values  normalized.

In Sect.~\ref{sub:coeffs}, IPC coefficients will be derived for one of the 16 flight detectors that will serve as a reference for all the studies detailed hereafter. In the following subsections, a method of correcting IPC taking into account spatial variations will be described and validated using dedicated data taken at CPPM for characterization.

\subsection{Derivation of IPC coefficients}
\label{sub:coeffs}

Several experimental protocols have been used to characterize IPC, some through the illumination of a single pixel with $^{55}$Fe sources\cite{Fox_2009} or taking advantage of cosmic ray hits\cite{donlon_2017}, others using electrical charge generation in hot pixels\cite{Giardino_2012} or even through single pixel reset (SPR), a technique using an HxRG dedicated feature allowing the reset of a single pixel to a voltage different from its neighbors\cite{dudik_2012}, which are reset at 300mV. This latter technique is the only one that can actually decorrelate IPC from charge drifting and allows per pixel IPC coefficients estimation. This technique proposed by Seshadri et al.\cite{seshadri_2008} has been used to characterize \Euclid NISP detectors during the on-ground characterization campaign. 

Actually we modified the standard IPC/SPR analysis to better take into account channel common mode noise and the transient effect that appears after the reset. Our measurements show that coupling between neighboring pixels becomes negligible at the second-closest pixels. Therefore, the IPC kernel may be limited to a $3 \times 3$ pixel kernel.

The data used here consist of two SPR measurements acquired at a reset voltage of $500$\,mV corresponding to an approximately $60$\,ke accumulated signal well below the full well ($\simeq$ 137\,ke). Each measurement is made up of 16 frames before the reset and 16 after that were averaged to reduce noise component. The two measurements were also averaged to improve accuracy. Uncertainties on the coefficients were estimated by propagation of uncertainties coming from the measured ADU signal and evaluated to about 0.01\,\%.
No pixel was masked so that cosmic ray hits will impact our results by adding charge drifting to IPC. An algorithm is under development to take into account cosmic rays impact on IPC.

\begin{figure} [ht]
		\begin{center}
			\begin{tabular}{c} 
				\includegraphics[width=.9\linewidth]{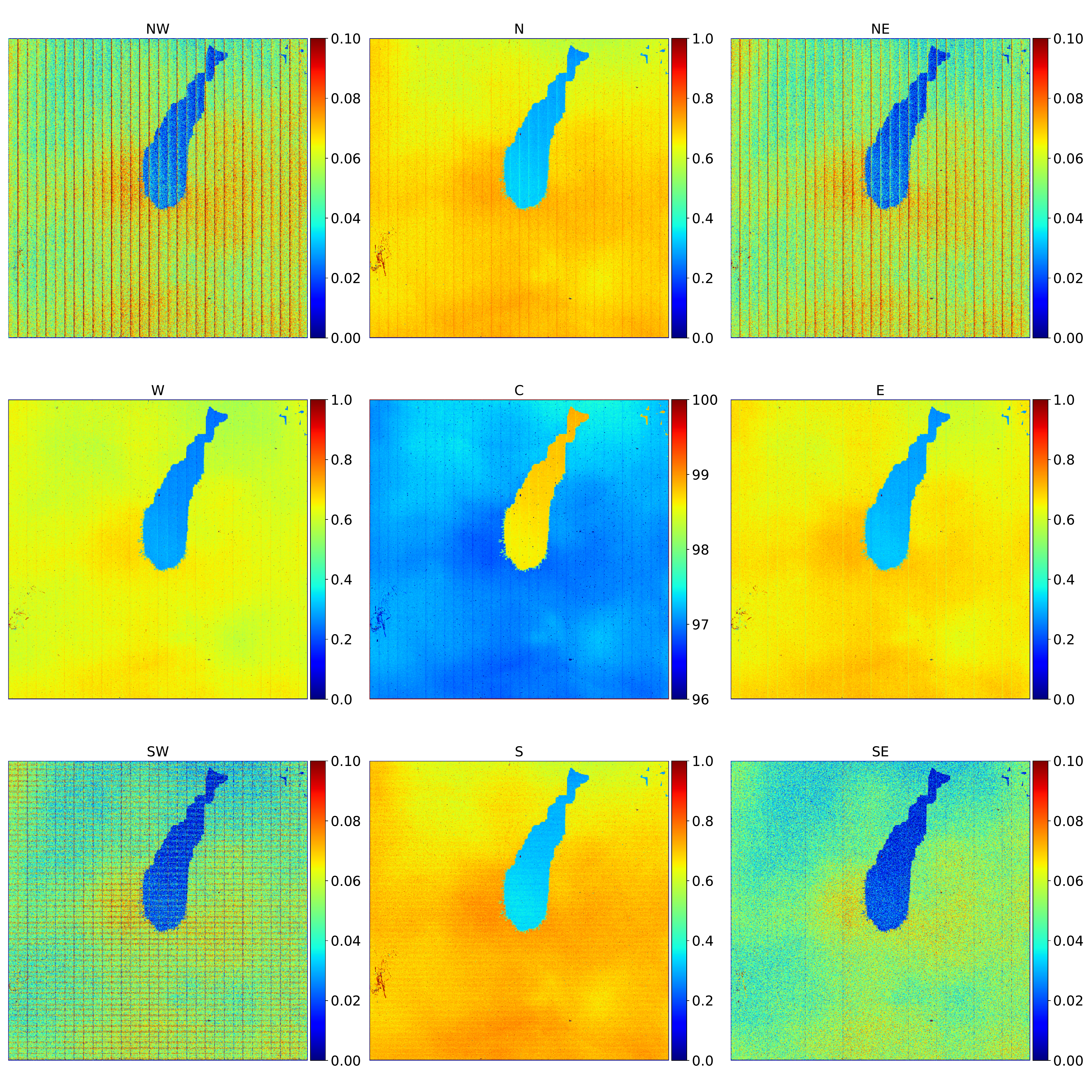} 
			\end{tabular}
		\end{center}
		\caption[example] 
		{ \label{fig:spr}  
	    Maps of IPC coefficients for a $3 \times 3$ kernel and a 500\,mV bias measured via SPR method. Coefficients correspond to the percentage of signal detected on a pixel that originates from the central pixel. Central pixel looses about 2.87 $\pm$ 0.01\,\% (median over the detector) signal due to IPC. Effect of IPC on the signal of epoxy void (yellow for C map) is 120\,\% lower than that in epoxy filled region (blue for C map).}
	\end{figure}

The results of our measurements are presented in Fig.~\ref{fig:spr} as 9 $2040 \times 2040$ maps labeled with cardinal directions. Each map shows, for all the pixels, the percentage of signal detected on the neighboring pixel corresponding to its cardinal position with respect to the central reset pixel. For example the values of the west map are the percentage of signal that will be detected on the left neighbors of all the pixels. Maps NW, NE, SW and SE will also be referred to as diagonals.

From Fig.~\ref{fig:spr} it may be observed that the central pixel loses about 2.87 $\pm$ 0.01\,\% (median over the detector) signal due to IPC. While several studies\cite{Giardino_2012, Fox_2009} use a simple kernel with a single $\alpha$ value for N, E, S, W cardinal directions and 0 for the diagonals of the kernel, here median IPC coefficients of the 4 cardinal neighbors vary by 10\,\% and should thus be treated as 4 different coefficients. These variations are possibly due to the readout direction of the matrix. Likewise, considering the uncertainty on IPC coefficients and the background noise level, coupling with the diagonals NW, NE, SW and SE may not be neglected and non-zero values should be retained. Moreover, differences larger than 10\,\% are also visible between the median values of diagonals. Therefore, a full non-zero kernel of IPC coefficients will be supplied to the Euclid Consortium for each pixel of the detector.

\paragraph{Spatial variations}
Looking at the central pixel map of Fig.~\ref{fig:spr}, pixels may be categorized in two groups: the pixels at the center of the detector (in yellow) and the rest of the pixels (in blue). Clearly, if IPC coefficients within those two areas vary little (see below), a large difference appears between the coefficients of the two areas. Previous studies have flagged this ``yellow'' area as an epoxy void\cite{Rauscher_2015}, i.e., an area where the space between the indium bumps is not filled with epoxy resin. For our reference detector,  IPC coefficients of this epoxy void zone are a 120\,\% smaller than that of the ``blue'' part of the detector, which is actually filled with epoxy. This suggests that the epoxy layer is responsible for more than a half of the IPC effect. This observation is consistent with the theory suggested by several studies\cite{donlon_2017, Kannawadi_2016, moore_interpixel_2004} that IPC mostly comes from fringing fields between the edges of the pixel implants. Due to the dielectric constant of epoxy, the coupling capacitance formed is greater with epoxy than without, so that corresponding IPC is larger. The dielectric constant of pure epoxy resin has been reported to be about 3.5\cite{wang_simultaneously_2021}, which means coupling capacitance of epoxy filled regions should be 250\,\% higher than epoxy void ones. Given that coupling cannot occur in the depleted layer, if the dielectric constant of the epoxy used by Teledyne is greater than 2.2, then a part of IPC should come from the multiplexer.

\FloatBarrier

As mentioned earlier, in addition to the difference between the two regions, variations higher than 20\,\% may also be observed within each region regardless of the epoxy filling. Finally, pixels at the boundaries between channels appear to have different IPC coefficients but not at the same level for all the directions. Our hypothesis is that this effect could come from the vertical reset control line that affects the signal in the first column of channels.
To account for these spatial fluctuations in the evaluation of detector performance, a per pixel IPC definition and correction is mandatory.

\medskip
\paragraph{Influence of signal level on the IPC coefficients}
Different values of the SPR reset voltage have been tested in order to investigate the impact of signal level on the IPC coefficients. Voltages 375\,mV, 500\,mV and 625\,mV were applied corresponding to pixel accumulated signal states of respectively 37\,ke, 60\,ke and 80\,ke (considering a median baseline of 30\,ke). The central IPC coefficient corresponding to the percentage of signal detected on the central pixel and its associated error are compiled in table~\ref{tab:vreset} for the three applied voltages. From this table, it appears that the IPC coupling measured by the SPR method is basically constant over a large dynamic range. These results are consistent with other measurements from HxRG detectors obtained with hot pixels\cite{Giardino_2012} or SPR\cite{dudik_2012}.

	\begin{table}[ht!]
		\caption{Median central IPC coefficient for 3 different reset voltages}
		\label{tab:vreset}
		\begin{center}       
			\begin{tabular}{|l|c|c|c|}
				\hline
				\rule[-1ex]{0pt}{3.5ex}  Vreset & 375\,mV & 500\,mV & 625\,mV   \\
				\hline 
				\rule[-1ex]{0pt}{3.5ex}  Equivalent signal level & 37\,ke & 60\,ke & 80\,ke   \\
				\hline 
				\rule[-1ex]{0pt}{3.5ex}  IPC coefficient [median $\pm$ uncertainty] & 97.12\,$\pm$\,0.04\%  &  97.13\,$\pm$\,0.01\% & 97.14\,$\pm$\,0.01\%   \\
				\hline 
			\end{tabular}\\
		\end{center}
	\end{table}
	
\FloatBarrier

Classical IPC modeling defines the coupling coefficient as the value of pixel capacitance relative to coupling capacitance. Since pixel capacitance is known to be nonlinear and to increase with signal, the corresponding coupling coefficient is expected to increase with signal. However most papers using the SPR method report IPC coefficients constant with signal level, as in here. This constant behavior may be understood if the coupling capacitance increases similarly with the signal, so that the ratio remains approximately constant. 


Other studies report an increase of cross-talk with signal. These are based on methods using $^{55}$Fe\cite{dudik_2012}, cosmic rays\cite{donlon_2017}, or flat field cross correlation\cite{Giardino_2012}, all of which measure an overall cross-talk and do not allow the distinction between IPC contribution and charge diffusion. The observed increase of cross-talk could thus simply be due to charge diffusion rather than IPC.

\subsection{IPC correction method}
\label{sub:deconv}

IPC alters the pixels' signal of each data frame at a different level for each pixel. The measurement of a detector's performance is based on analyzing the ramps of signal acquired during the characterization. Consequently to derive IPC-free detectors performance, a per pixel IPC correction using the SPR measurements is needed.

Donlon et al. \cite{donlon_2018} developed an iterative decoupling algorithm that provides a per pixel IPC correction, but the computing time is way too large considering the enormous amount of data to be treated. A faster correction may be achieved thanks to the Fourier space deconvolution proposed by McCullough et al. \cite{mccullough_2008}. Hereafter the latter method is presented with some adaptation to achieve a per pixel IPC correction.

As introduced earlier, IPC is typically modeled by a $3 \times 3$ convolution kernel as described by equation~\ref{eq:conv_kern}. The SPR method provides the values of $\alpha_i$ for all $2040 \times 2040$ ``science" pixels of our detector, basically a reformated version of IPC coefficients derived in Sect.~\ref{sub:coeffs}. 

\begin{equation}
k =  \begin{bmatrix} 
    \alpha_1 & \alpha_2 & \alpha_3 \\
    \alpha_4 & 1 - \sum \limits_{i=1}^8 \alpha_i & \alpha_5 \\
    \alpha_6 & \alpha_7 & \alpha_8 
    \end{bmatrix} 
    \label{eq:conv_kern}
\end{equation}

\FloatBarrier

In terms of IPC (putting aside any other effect), a measured frame including IPC may be modeled as an IPC-free frame convolved by an IPC kernel $k$, as introduced in Eq.~\ref{eq:conv_kern}. The key to McCullough's method is to find a kernel $g$ whose convolution with kernel $k$ gives a $\delta$-function. Then, convolving the measured frame by kernel g will yield an IPC-free frame. Since, here, our goal is to deconvolve data per pixel, McCullough's algorithm needs to be modified so as to compute an individual $g$ kernel for each pixel. This may be achieved using the Eq.~\ref{eq:compute_g} where $F$ describes the Fourier transform. 

\begin{equation}
\label{eq:compute_g}
    g =  F^{-1} \left( \dfrac{F \left(\delta \right)}{F\left( k \right)} \right) \quad with \quad \delta = \begin{bmatrix} 
    0 & 0 & 0 \\
    0 & 1 & 0 \\
    0 & 0 & 0 \\
    \end{bmatrix}
\end{equation}

Equation~\ref{eq:compute_g} provides a deconvolution kernel for all the pixels of the detector which will be used to rebuild the true signal $S_{i,j}$ of a pixel according to the signal of its neighbors using Eq.~\ref{eq:deconv}. In this equation, $i$ and $j$ are the row and column indices of the pixel and $S'$ is the measured signal of a pixel. The signal of the neighbors are also affected by coupling with their own neighbors and this is why we used a $5 \times 5$ kernel rather than a $3 \times 3$ one.

\begin{equation}
\label{eq:deconv}
     S_{i,j} =   
    \begin{bmatrix} 
    S'_{i-2,j-2} & S'_{i-2,j-1} & S'_{i-2,j} & S'_{i-2,j+1} & S'_{i-2,j+2} \\
    S'_{i-1,j-2} & S'_{i-1,j-1} & S'_{i-1,j} & S'_{i-1,j+1} & S'_{i-1,j+2} \\
    S'_{i,j-2} & S'_{i,j-1} & S'_{i,j} & S'_{i,j+1} & S'_{i,j+2} \\
    S'_{i+1,j-2} & S'_{i+1,j-1} & S'_{i+1,j} & S'_{i+1,j+1} & S'_{i+1,j+2} \\
    S'_{i+2,j-2} & S'_{i+2,j-1} & S'_{i+2,j} & S'_{i+2,j+1} & S'_{i+2,j+2} \\
    \end{bmatrix} 
    *
    g_{i,j}
\end{equation}

Since the kernel $g$ is computed based on the IPC of the central pixel, the IPC effect from the pixel $[i,j+2]$ on the pixel $[i,j+1]$ is approximated to be equal to the effect of the pixel $[i,j+1]$  on the central pixel. This approximation is acceptable considering that IPC variations are very low between two neighboring pixels. However, in regions with strong variations, such as at the epoxy void boundary or close to reference pixel lines and columns, processing needs to be adapted.

\subsection{Validation with characterization data}
\label{sec:valid_deconv}
\begin{figure} [ht!]
		\begin{center}
			\begin{tabular}{c} 
				\includegraphics[height=5.6cm]{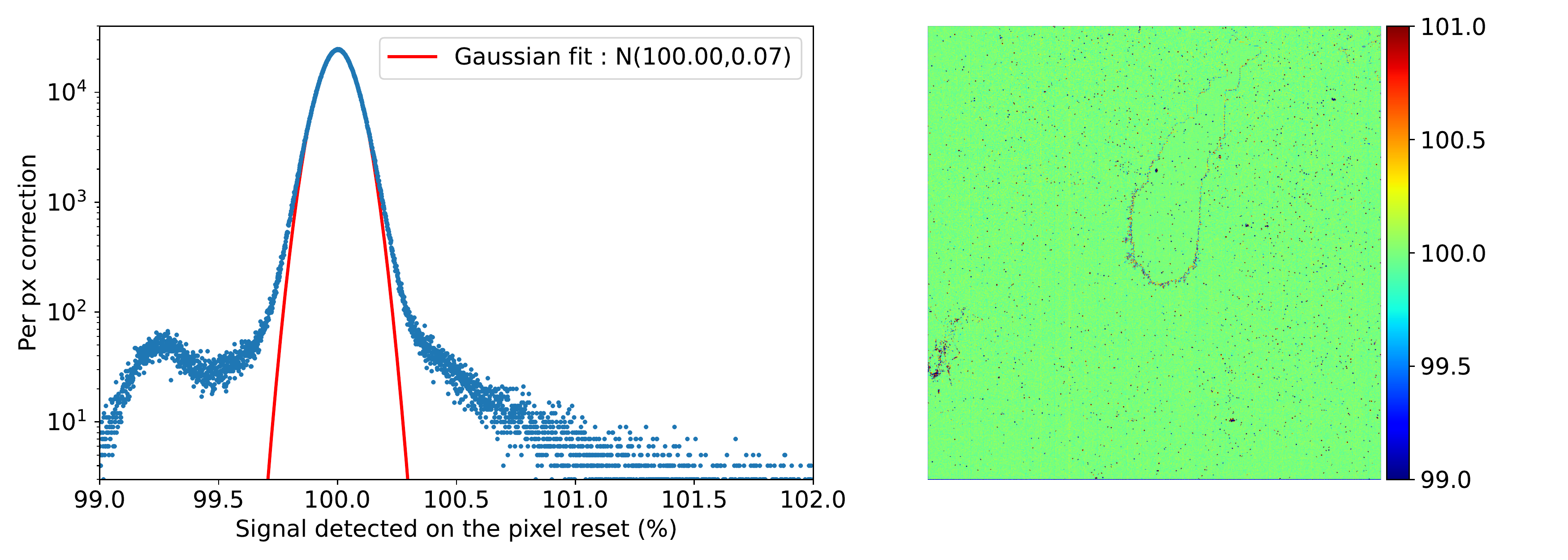}
			\end{tabular}
		\end{center}
		\caption[example] 
		{ \label{fig:clearipc_px}  
		Histogram (left) and map (right) of the percentage of signal detected on the central pixel after per pixel IPC correction. A normal law may be fitted to the main peak with a mean value of 100\%, which means the correction is accurate. The peak at 99.2\,\% comes from edge effects which will be taken into account in the future.  }
	\end{figure}

In order to validate the method described above, two sets of SPR data taken in the same conditions were used. One set was used to derive the IPC correction, which was then applied to the other set. In case of an ideal deconvolution of IPC, we expect a Gaussian distribution of the reconstructed pixel value, with a mean of 100\,\%. The results of the deconvolution are presented in Fig.~\ref{fig:clearipc_px} through a histogram (left) and a map (right). As expected, a normal law may clearly be fitted to the main peak of the histogram giving a mean value of 100\,\% and a standard deviation of 0.07\,\% that demonstrate the reliability of our method. The smaller peak at 99.2\,\% has been checked to come from the pixels on the boundaries of the epoxy void or from the reference pixel lines/columns. This peak is clearly due to the method, which should obviously be adapted in order to better take into account these boundaries. Hereafter, all the pixels not included in the $5-sigma$ confidence interval of the Gaussian fit will be masked: they represent less than 1.5\,\% of all pixels.

\FloatBarrier

\begin{figure} [ht!]
		\begin{center}
			\begin{tabular}{c} 
				\includegraphics[height=5.6cm]{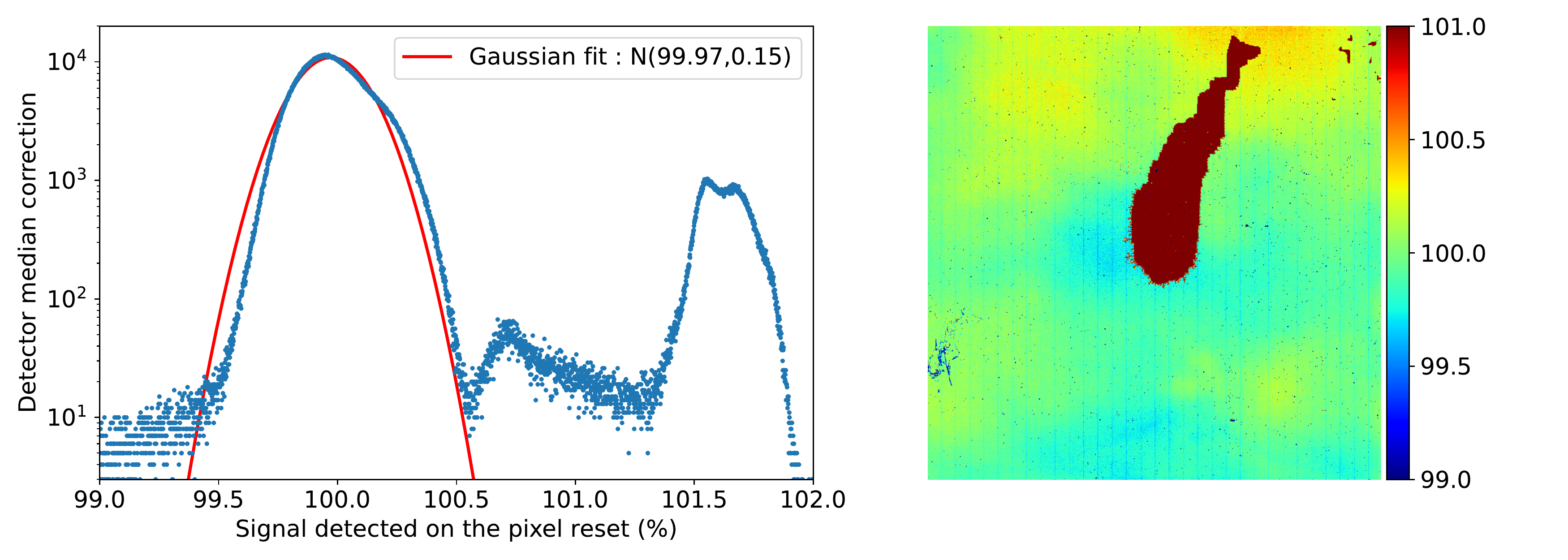}
			\end{tabular}
		\end{center}
		\caption[example] 
		{ \label{fig:clearipc_det}  
		Histogram (left) and map (right) of the percentage of signal detected on the central pixel after IPC correction using a median IPC kernel. The epoxy void region is overcorrected creating a 1.5\,\% bias on the signal.}
	\end{figure}

As a mean to show the necessity of a per pixel IPC correction, two corrections are compared here: one with individual pixel kernels and another with a median kernel for all pixels. While Fig.~\ref{fig:clearipc_px} shows a per pixel kernel correction, Fig.~\ref{fig:clearipc_det} shows similar graphs for an overall median kernel. 

\FloatBarrier

In Fig.~\ref{fig:clearipc_det}, the histogram may still be fitted by a Gaussian of mean close to 100\,\%, but the standard deviation is doubled. The peak near 101.5\,\% corresponds to pixels within the epoxy void area. Clearly these are overcorrected creating a systematic bias higher than 1.5\,\% on the corrected signal. Another possibility is to define only two different kernels by detector, one for the epoxy void and one for the region filled with epoxy. 
The actual number of kernels chosen to describe IPC's spatial variations will depend on the accuracy needed by the mission. Since this study aims at probing the consequences of physical spatial variations on the detector's performance, in the following a per pixel IPC correction will be used.

\section{CONVERSION GAIN}
\label{sec:conversiongain}
One important parameter measured during characterization is conversion gain (ADU\,e$^{-1}$). It is defined as the numbers of ADUs that represent an electron. It is all the more critical to properly evaluate the conversion gain than QE, dark current, and read noise, three other fundamental parameters, are calculated using previously derived conversion gain values so that the accuracy obtained for these three parameters directly depends on that of conversion gain. Beside the capacitance comparison technique~\cite{finger_conversion_2005-1} that requires an external capacitance, conversion gain is usually derived through the ``photon transfer curve'' (PTC) defined by Janesik et al.\cite{janesick_scientific_2001} or by ``mean variance"  method defined by Mortara et al.\cite{Mortara_1981}.

This last method takes advantage of the Poisson statistics of incident photons to measure the conversion gain as the slope of the variance vs. mean curve under the assumptions of a negligible gain variance and a linear response of the pixels. In this paper, this method is referred to as the temporal version of mean variance because variance and mean are estimated over multiple realizations of the same pixel at different times. Because of the limited time dedicated to characterization (45 days for each flight detector), a temporal mean variance method did not allow us to reach the accuracy level of $1$\,\% maximum relative uncertainty we want to achieve. Therefore a compromise needs to be obtained by combining temporal and spatial statistics.

The adapted method which we refer to as spatial version of mean variance was described by Secroun et al.\cite{secroun:hal-01959780}. It estimates the variance and mean over a N$\times$N super-pixel and averages the results over multiple realizations taken at different times. This method assumes that spatial variance is equivalent to temporal variance. This assumption cannot be satisfied in the presence of interpixel capacitance because it creates spatial correlations that reduce spatial variance. Consequently, in order to use the spatial mean-variance method, the IPC effect needs to be taken into account. Moore et al.~\cite{moore_quantum_2006} proposed to add to the spatial variance calculation the correlations terms between neighboring pixels that include the IPC effect. Other studies introduced an analytical coefficient that can correct the IPC impact on conversion gain\cite{mccullough_quantum_2008,smadja:in2p3-00467907}.
Recently, Hirata et al.~\cite{hirata_2019} developed a method that uses spatial and temporal correlations in flat fields to derive conversion gain among other parameters. All these methods do not use a priori knowledge of IPC coefficients but only flat field data to derive conversion gain. 

Hereafter, IPC correction is used to derive an IPC-free conversion gain at the pixel level, which will be compared to the conversion gain obtained thanks to Moore's method. Hirata's method is not considered hereafter because its implementation is not straightforward. Nor are analytical approaches which correct IPC at the superpixel level preventing a per pixel correction.

\subsection{Methodology}
\label{sub:cg_method}
The spatial mean variance method estimates variance and signal over a super-pixel made of N\,x\,N pixels. Since calculations are done on spatially distributed pixels, fixed pattern noise (FPN) must be removed by evaluating variance on the difference of two ramps, frame to frame following Eq.~\ref{eq:mean_var}.

\begin{equation}
\label{eq:mean_var}
    \mathrm{Mean} = \dfrac{1}{N^2} \sum \limits_{i,j=0}^{N} \dfrac{S_1[i,j] + S_2[i,j]}{2} \quad ; \quad \mathrm{Var} = \dfrac{1}{N^2-1} \sum \limits_{i,j=0}^{N} \dfrac{(S_1[i,j] - S_2[i,j])^2}{2}
\end{equation}

One value of conversion gain may thus be derived from a single pair of ramps. Conversion gain is then estimated as the mean of all conversion gain values calculated from pairs of ramps. The uncertainty on this mean conversion gain is defined as the uncertainty on the mean value, namely $\sigma / \sqrt{\mathrm{M}}$ where M is the total number of calculated conversion gain values. All the ramps we used were made up of 394 frames read every 1.45 seconds.

In order to properly apply spatial mean variance and avoid discrepancies in the calculations, outlier pixels such as dead or hot pixels must be removed or masked. Thus, several masks were applied that remove overall less than 2\,\% of the entire matrix, outlined below:
\begin{itemize}
    \item Mask on  disconnected pixels;
    \item Mask on pixels defined in Sect.~\ref{sec:valid_deconv} that are not properly corrected from IPC;
    \item Mask on pixels with a high baseline to avoid nonlinear regime and saturation;
    \item Mask on highly nonlinear pixels thanks to a quality factor based on the goodness of a linear fit on the ramp\cite{kubik:in2p3-01388431}.
\end{itemize}

Since the mean variance method assumes a linear response of the pixels, the pixel nonlinear regime must be avoided. For this purpose, only ramps acquired with an input flux lower than 90\,e\,s$^{-1}$, which corresponds to 60\,\% of the full well at maximum, are used.

Moreover, it has been shown in a previous work~\cite{secroun:hal-01959780} that persistence can heavily affect the apparent flux, especially in the first frames of a ramp, by for instance trapping charges, which in turn reduces the ADU signal. Ignoring the 100 first frames of the ramps will mitigate the persistence effect. 

Finally, because persistence shows a long time decay component, two ramps acquired with the same parameters could be slightly different depending on the previous illumination. Consequently the mean of the differences of the two ramps will not be null and Eq.~\ref{eq:mean_var} used by standard mean variance method to estimate variance will be biased. To handle this issue, the variance estimator has been changed to Eq.~\ref{eq:var_lat} where $D[i,j]$ is the difference frame and $\bar{D}[i,j]$ its mean over the super-pixel.

\begin{equation}
\label{eq:var_lat}
    \mathrm{Var} = \dfrac{1}{N^2-1} \sum \limits_{i,j=0}^{N} \dfrac{(D[i,j] - \Bar{D}[i,j])^2}{2} \quad \textrm{with} \quad D[i,j] = S_1[i,j] - S_2[i,j]
\end{equation}

In summary, 521 pairs of ramps fulfill all our selection criteria. The estimated conversion gain will be presented in the next section.

\subsection{Per superpixel IPC-free conversion gain}
\label{sub:cgmap}

\begin{figure} [ht!]
		\begin{center}
			\begin{tabular}{c} 
				\includegraphics[height=5.6cm]{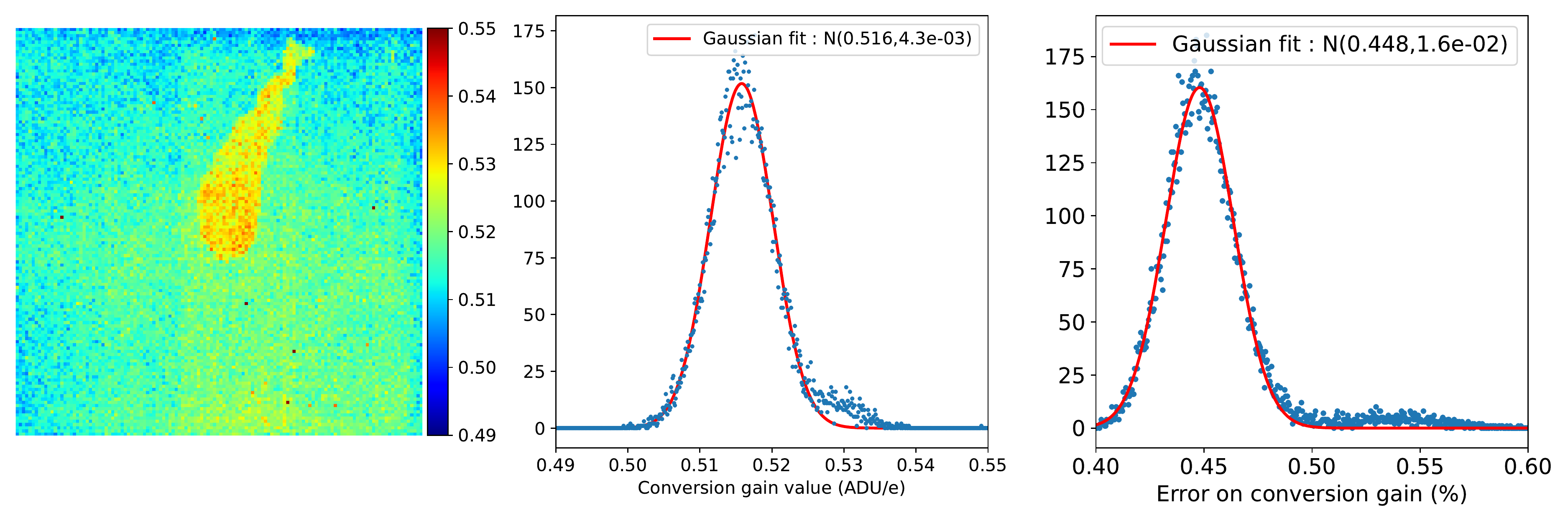}
			\end{tabular}
		\end{center}
		\caption[example] 
		{ \label{fig:gainpx16}  
		Map of conversion gain (left) per 16x16 superpixel estimated after IPC correction, the corresponding histogram (middle) and the histogram of the error on the conversion gain (right). The mean gain in the epoxy filled area is $0.515 \pm 0.002$\,ADU\,e$^{-1}$ while that in the unfilled area is $0.528 \pm 0.002$\,ADU\,e$^{-1}$.}
	\end{figure}

The results of the conversion gain calculations are shown in Fig.~\ref{fig:gainpx16} with: from left to right, a per super-pixel of size 16$\times$16 IPC-free conversion gain map, the corresponding histogram and the histogram of the error on our per super-pixel gain estimator. The mean value of estimated conversion gain over the detector is $0.516 \pm 0.002$\,ADU\,e$^{-1}$. Given the selected data, a super-pixel of 8$\times$8 pixels is our limit to keep an uncertainty on the estimated gain inferior to 1\%. 

\FloatBarrier

Looking at Fig.~\ref{fig:gainpx16}, the same two regions noticed in Fig.~\ref{fig:spr} are still visible, despite the applied IPC correction. It means that the epoxy layer does not only affect  the IPC but also the measure of the conversion gain. However, the mean gain in the epoxy filled area is $0.515 \pm 0.002$\,ADU\,e$^{-1}$ while that in the unfilled area is $0.528 \pm 0.002$\,ADU\,e$^{-1}$, namely a 2.2\% difference in the conversion gain, far below the difference in IPC effect. 
Other identifiable regions showing spatial variations are the 32 readout channels. Because all pixels of a channel share a common source follower (buffer amplifier), the fraction of the conversion gain that comes from this source follower is equal for all these pixels. All in all, variations higher than 1\% are visible within the epoxy filled region.
This means that taking a mean value over the entire detector for the conversion gain causes an error that can be greater than 2\%. Since the conversion gain is used to derive read noise, dark current, and quantum efficiency, this error will propagate to those performance parameters. 
With a view to mitigate the effect of spatial variations on the detector's performance, for the \Euclid mission, the instrument development team chooses to define two values of conversion gain per readout channel, one for the epoxy filled region and one for the epoxy void region.

\subsection{Comparison with other methods}
\label{sub:comp}

\begin{figure} [ht!]
		\begin{center}
			\begin{tabular}{c} 
				\includegraphics[height=6cm]{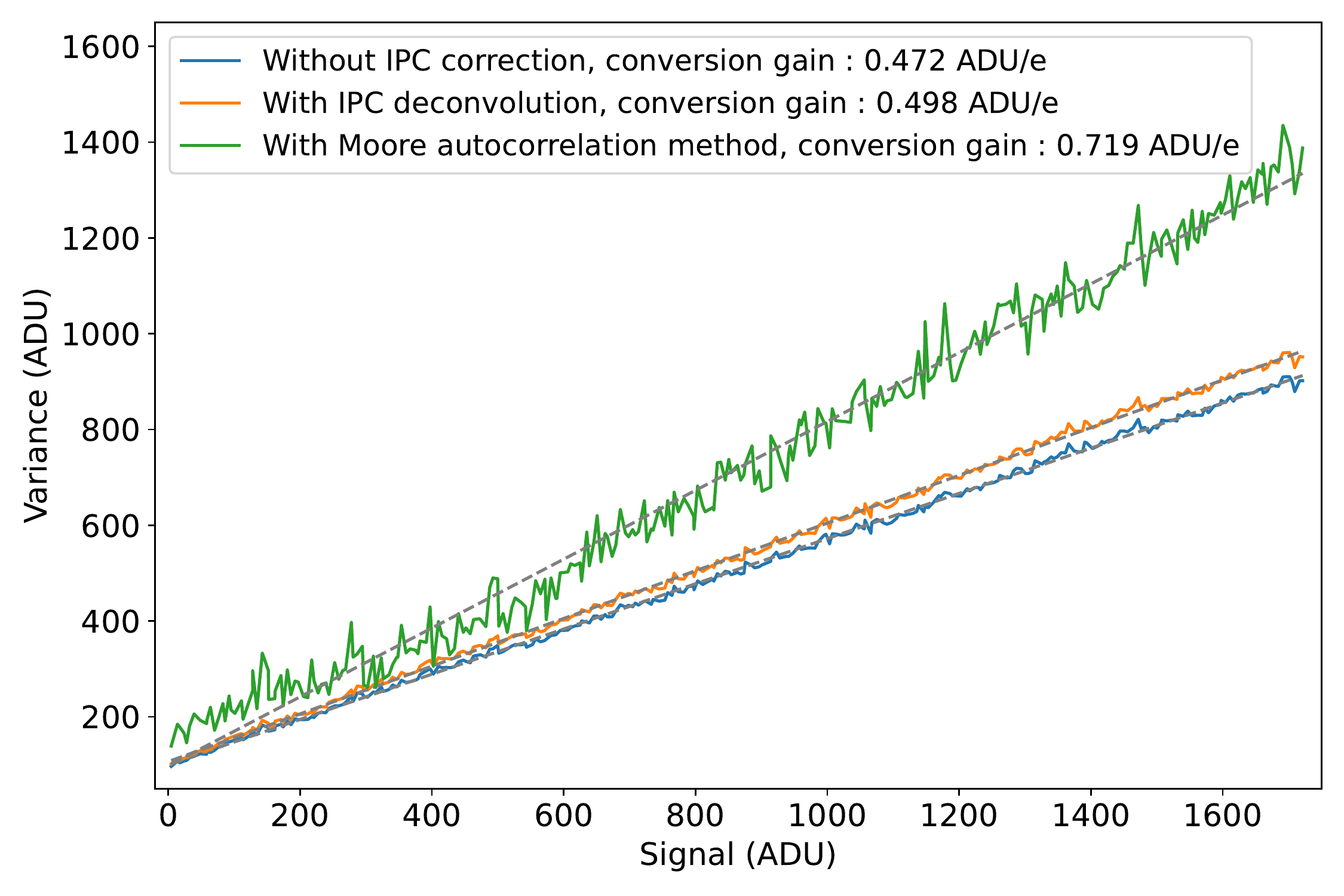}
			\end{tabular}
		\end{center}
		\caption[example] 
		{ \label{fig:compptc}  
		Variance vs. signal curves for 3 cases: spatial mean variance gain, IPC corrected gain, and Moore's gain. Variance and signal are estimated from a pair of 8 e\,s ramps on a 64x64 super-pixel. Moore's method overcorrects the estimated variance as well as adds a substantial quantity of noise.}
	\end{figure}

Correcting IPC in individual frames allowed us to calculate a conversion gain free from IPC effect. But this correction has a cost in computing time and data storage. Existing methods without IPC a priori such as Moore's method are less time-consuming, so it is of interest to compare their reliability to our approach. Figure~\ref{fig:compptc} shows variance vs. signal curve calculated from raw data (blue), IPC corrected data (red) and using Moore's method (green) on a 64$\times$64 super-pixel ramp with an $8$\,e\,s$^{-1}$ input flux. 

The two variance vs. signal curves calculated with raw and IPC corrected data show a similar behavior. However, as is expected, due to the presence of IPC, the estimated variance from raw data is lower than that with IPC corrected data. As for Moore's method, it clearly overcorrects the estimated variance as well as adds a substantial quantity of noise. The issue with Moore's method is that it interprets all spatial correlations as coming from IPC even when not. Indeed these correlations are expected to also contain contribution from charge diffusion or persistence even if limited by our data selection.

\FloatBarrier

\begin{figure} [ht!]
		\begin{center}
			\begin{tabular}{c} 
				\includegraphics[height=6cm]{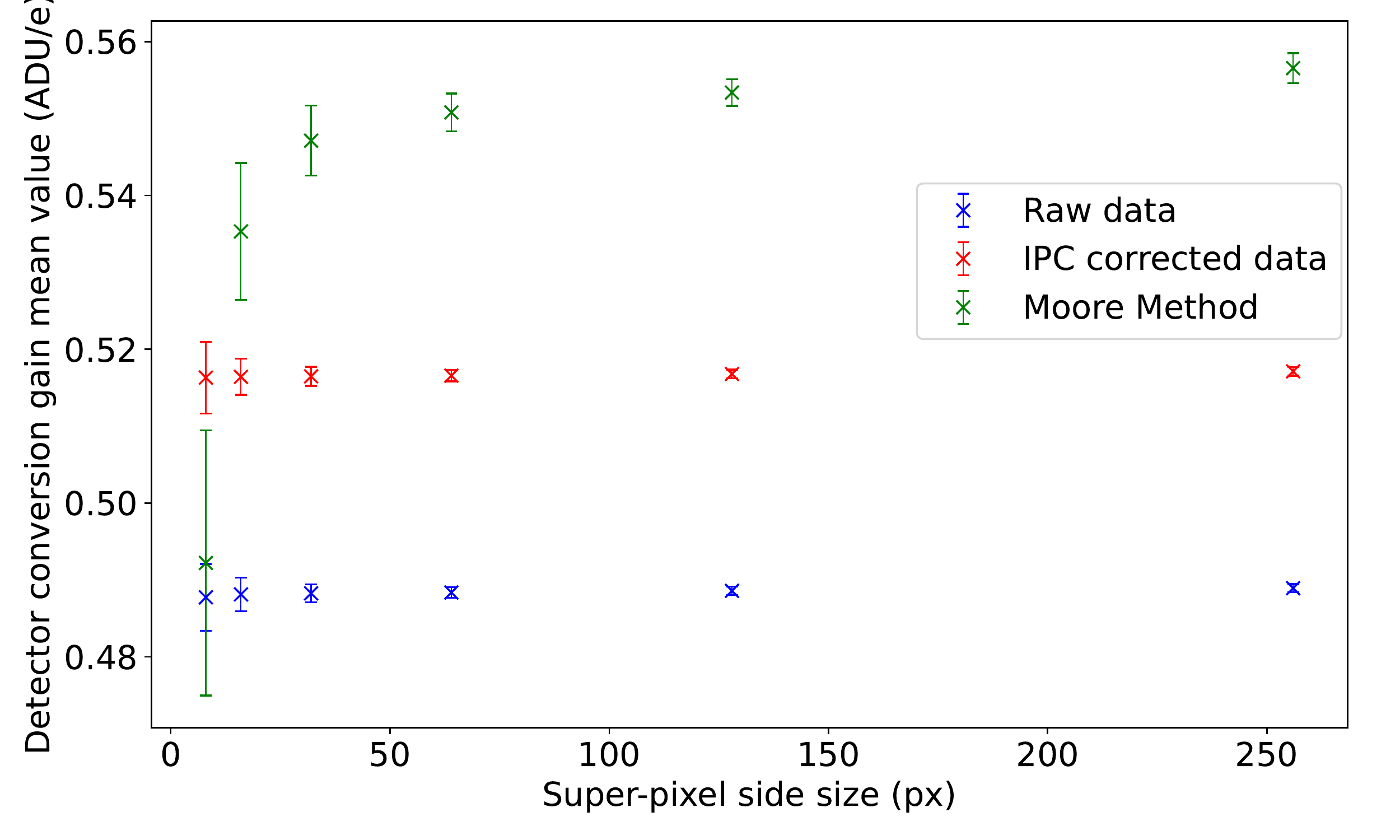}
			\end{tabular}
		\end{center}
		\caption[example] 
		{ \label{fig:comppxsize}  
		Conversion gain as a function of super-pixel size for 3 cases: spatial mean variance gain, IPC corrected gain, and Moore's gain. Gain estimated from IPC-free data is 5.6\% higher than that from raw data. IPC impacts the measured conversion gain twice more than the signal. }
	\end{figure}

Following the steps of several studies that have investigated the impact of super-pixel size on the gain measurements\cite{smadja:in2p3-00467907, mccullough_quantum_2008}, Fig.~\ref{fig:comppxsize} plots the detector average conversion gain as a function of super-pixel size for the three previously considered cases. The conversion gains are calculated in the same way as in Sect.~\ref{sub:cg_method} and averaged over the detector.  Error bars corresponding to the mean uncertainty on a super-pixel gain are also shown. This figure shows that, with the spatial mean variance method defined above, the size of the superpixel has no impact on the conversion gain for neither the raw data nor the IPC corrected data unlike in McCullough's measurements\cite{mccullough_quantum_2008} and for the gain estimated by Moore's method which decreases with pixel size.

Besides, Fig.~\ref{fig:comppxsize} exhibits an underestimation of approximately 5.6\,\% of the gain in the raw data case compared to IPC corrected data. This is consistent with the estimation given by Moore\cite{moore_interpixel_2004}, which states that the effect of IPC on the measurement of gain is twice its effect on signal measurement. Finally, Moore's method is not adapted to our detectors that clearly exhibit other sources of spatial correlations than IPC, typically persistence that was mentioned earlier. 

\FloatBarrier
\section{CONCLUSION}
\label{sec:conclu}
The spatial variations of a flight \Euclid H2RG interpixel capacitance and conversion gain have been explored  using on ground characterization data acquired at CPPM. 

Thanks to SPR method, per pixel IPC effect was measured on the first neighboring pixel ring and a median loss of signal of about $2.87 \pm 0.01$\,\% on the central pixel was observed. This loss of signal presents spatial variations greater than 120\,\%, which have been flagged to originate from epoxy void between sensitive layer and multiplexer. Those measurements also show spatial variations higher than 20\,\% within the epoxy filled region. Furthermore, our results exhibit IPC anisotropy for our detector.

These SPR measurements lead  to the definition of a new method based on McCullough's\cite{mccullough_2008} that corrects IPC on individual frames at the pixel level. The method was validated using two sets of SPR data taken in the same condition. Comparing our results with the same method using a mean value of IPC over the detector exhibits a bias of 1.5\,\% on the corrected signal if using mean value.

IPC-free conversion gain was derived using mean variance method with per pixel IPC corrected data. A spatial version of mean variance method in addition to a proper selection of data resulted in an error on the conversion gain of 16x16 super-pixels inferior to 1\,\%. Results of our measure show that the conversion gain of pixels within the epoxy void region is 2.2\,\% higher than that in the epoxy filled region meaning that epoxy also affects the derivation of conversion gain. In addition spatial variations in the epoxy filled region are higher than 1\,\%. We conclude that taking into account these variations will greatly enhance the detector performance's evaluation.

The method developed by Moore\cite{moore_quantum_2006} to derive conversion gain taking IPC into account was implemented and compared to ours to verify its reliability. Adding a neighboring pixel correlation term to estimate IPC's impact seems to be impossible with our detectors because of all the spatial correlations created by, for instance, charge diffusion or persistence. 

Beyond IPC, other parameters such as nonlinearity and persistence are expected to show spatial variations and correlations and should be explored. Depending on their impact, we might need to adapt our methods further to take them into account and derive a conversion gain free of nonlinearity and persistence.


\acknowledgments 

This work was developed within the frame of a CNES-CNRS funded Phd thesis. 
 
\AckECon

\bibliography{report} 

\begin{thebibliography}{10}

\bibitem{Racca_2016}
Racca, G.~D. et~al., ``{The Euclid mission design},'' in [{\em Space Telescopes
  and Instrumentation 2016: Optical, Infrared, and Millimeter
  Wave}{\nolinebreak\hspace{0.1em}]},  MacEwen, H.~A., Fazio, G.~G., Lystrup,
  M., Batalha, N., Siegler, N., and Tong, E.~C., eds.,  {\bf 9904},  235 --
  257, International Society for Optics and Photonics, SPIE (2016).

\bibitem{Maciaszek2016}
Maciaszek, T. et~al., ``Euclid near infrared spectrometer and photometer
  instrument concept and first test results obtained for different breadboards
  models at the end of phase c,'' in [{\em Space Telescopes and
  Instrumentation}{\nolinebreak\hspace{0.1em}]},  {\em Proc. of SPIE} {\bf
  9904},  99040T (2016).

\bibitem{JWST}
Birkmann, S.~M., Ferruit, P., Rawle, T., Sirianni, M., de~Oliveira, C.~A.,
  Böker, T., Giardino, G., Lützgendorf, N., Marston, A., Stuhlinger, M.,
  te~Plate, M. B.~J., Jensen, P., Rumler, P., Dorner, B., Karl, H., Mosner, P.,
  Wright, R.~H., and Rapp, R., ``{The JWST/NIRSpec instrument: update on status
  and performances},'' in [{\em Space Telescopes and Instrumentation 2016:
  Optical, Infrared, and Millimeter Wave}{\nolinebreak\hspace{0.1em}]},
  MacEwen, H.~A., Fazio, G.~G., Lystrup, M., Batalha, N., Siegler, N., and
  Tong, E.~C., eds.,  {\bf 9904},  92 -- 102, International Society for Optics
  and Photonics, SPIE (2016).

\bibitem{WFIRST}
Gong, Q., Content, D.~A., Dominguez, M., Emmett, T., Griesmann, U., Hagopian,
  J., Kruk, J., Marx, C., Pasquale, B., Wallace, T., and Whipple, A.,
  ``{Wide-Field InfraRed Survey Telescope (WFIRST) slitless spectrometer:
  design, prototype, and results},'' in [{\em Space Telescopes and
  Instrumentation 2016: Optical, Infrared, and Millimeter
  Wave}{\nolinebreak\hspace{0.1em}]},  MacEwen, H.~A., Fazio, G.~G., Lystrup,
  M., Batalha, N., Siegler, N., and Tong, E.~C., eds.,  {\bf 9904},  403 --
  420, International Society for Optics and Photonics, SPIE (2016).

\bibitem{moons}
Cirasuolo, M. et~al., ``{MOONS: the Multi-Object Optical and Near-infrared
  Spectrograph for the VLT},'' in [{\em Ground-based and Airborne
  Instrumentation for Astronomy V}{\nolinebreak\hspace{0.1em}]},  Ramsay,
  S.~K., McLean, I.~S., and Takami, H., eds.,  {\bf 9147},  202 -- 214,
  International Society for Optics and Photonics, SPIE (2014).

\bibitem{Ariel}
{Tinetti}, G. et~al., ``{Ariel: Enabling planetary science across
  light-years},'' {\em arXiv e-prints} ,  arXiv:2104.04824 (Apr. 2021).

\bibitem{kubik_2014}
Kubik, B., Barbier, R., Castera, A., Chabanat, E., Ferriol, S., and Smadja, G.,
  ``{Impact of common modes correlations and time sampling on the total noise
  of a H2RG near-IR detector},'' in [{\em High Energy, Optical, and Infrared
  Detectors for Astronomy VI}{\nolinebreak\hspace{0.1em}]},  Holland, A.~D. and
  Beletic, J., eds.,  {\bf 9154},  427 -- 439, International Society for Optics
  and Photonics, SPIE (2014).

\bibitem{secroun_2016}
Secroun, A., Serra, B., Clémens, J.~C., Legras, R., Lagier, P., Niclas, M.,
  Caillat, L., Gillard, W., Tilquin, A., Ealet, A., Barbier, R., Ferriol, S.,
  Kubik, B., Smadja, G., Prieto, E., Maciaszek, T., and Sorensen, A.~N.,
  ``{Characterization of H2RG IR detectors for the Euclid NISP instrument},''
  in [{\em High Energy, Optical, and Infrared Detectors for Astronomy
  VII}{\nolinebreak\hspace{0.1em}]},  Holland, A.~D. and Beletic, J., eds.,
  {\bf 9915},  649 -- 657, International Society for Optics and Photonics, SPIE
  (2016).

\bibitem{plazas_2017}
Plazas, A., Shapiro, C., Smith, R., Rhodes, J., and Huff, E., ``Nonlinearity
  and pixel shifting effects in {HxRG} infrared detectors,'' {\em Journal of
  Instrumentation}~{\bf 12}(04),  C04009–C04009 (2017).

\bibitem{finger_conversion_2005-1}
Finger, G., Beletic, J.~W., Dorn, R., Meyer, M., Mehrgan, L., Moorwood, A.
  F.~M., and Stegmeier, J., ``Conversion {Gain} and {Interpixel} {Capacitance}
  of {CMOS} {Hybrid} {Focal} {Plane} {Arrays},'' {\em Experimental
  Astronomy}~{\bf 19}(1),  135--147 (2005).

\bibitem{Smith_2008}
Smith, R., Zavodny, M., Rahmer, G., and Bonati, M., ``A theory for image
  persistence in {HgCdTe} photodiodes,'' {\em Proceedings of SPIE - The
  International Society for Optical Engineering}  (2008).

\bibitem{moore_quantum_2006}
Moore, A.~C., Ninkov, Z., and Forrest, W.~J., ``Quantum efficiency
  overestimation and deterministic cross talk resulting from interpixel
  capacitance,'' {\em Optical Engineering}~{\bf 45}(7),  076402 (2006).
\newblock Publisher: International Society for Optics and Photonics.

\bibitem{barbier_2018}
Barbier, R., Ferriol, S., Kubik, B., Smadja, G., Secroun, A., Cl{\'e}mens,
  J.-C., Ealet, A., Gillard, W., Zoubian, J., Serra, B., Rosset, C., Kohley,
  R., Conversi, L., Fornari, F., and Buton, C., ``{Detector chain calibration
  strategy for the Euclid Flight IR H2RGs},'' in [{\em {SPIE Astronomical
  Telescopes + Instrumentation 2018}}{\nolinebreak\hspace{0.1em}]},   {\bf
  10709},  107090S (2018).

\bibitem{Kannawadi_2016}
Kannawadi, A., Shapiro, C.~A., Mandelbaum, R., Hirata, C.~M., Kruk, J.~W., and
  Rhodes, J.~D., ``The impact of interpixel capacitance in {CMOS} detectors on
  {PSF} shapes and implications for {WFIRST},'' {\em Publications of the
  Astronomical Society of the Pacific}~{\bf 128}(967),  095001 (2016).

\bibitem{Fox_2009}
Fox, O., Waczynski, A., Wen, Y., Foltz, R.~D., Hill, R.~J., Kimble, R.~A.,
  Malumuth, E., and Rauscher, B.~J., ``The $^{55}${Fe} {X-Ray} energy response
  of mercury cadmium telluride near-infrared detector arrays,'' {\em
  Publications of the Astronomical Society of the Pacific}~{\bf 121}(881),
  743--754 (2009).

\bibitem{donlon_2017}
Donlon, K., Ninkov, Z., Baum, S., and Cheng, L., ``Modeling of hybridized
  infrared arrays for characterization of interpixel capacitive coupling,''
  {\em Optical Engineering}~{\bf 56}(2),  024103 (2017).

\bibitem{Giardino_2012}
{Giardino}, G., {Sirianni}, M., {Birkmann}, S.~M., {Rauscher}, B.~J.,
  {Lindler}, D., {Boeker}, T., {Ferruit}, P., {De Marchi}, G., {Stuhlinger},
  M., {Jensen}, P., and {Strada}, P., ``{{NIRSpec} detectors: noise properties
  and the effect of signal dependent inter-pixel crosstalk},'' in [{\em High
  Energy, Optical, and Infrared Detectors for Astronomy
  V}{\nolinebreak\hspace{0.1em}]},  {Holland}, A.~D. and {Beletic}, J.~W.,
  eds., {\em Society of Photo-Optical Instrumentation Engineers (SPIE)
  Conference Series} {\bf 8453},  84531T (2012).

\bibitem{dudik_2012}
Dudik, R.~P., Jordan, M.~E., Dorland, B.~N., Veillette, D., Waczynski, A.,
  Lane, B.~F., Loose, M., Kan, E., Waterman, J., Rollins, C., and et~al.,
  ``Interpixel crosstalk in teledyne imaging sensors {H4RG-10} detectors,''
  {\em Applied Optics}~{\bf 51}(15),  2877 (2012).

\bibitem{seshadri_2008}
Seshadri, S., Cole, D.~M., Hancock, B.~R., and Smith, R.~M., ``{Mapping
  electrical crosstalk in pixelated sensor arrays},'' in [{\em High Energy,
  Optical, and Infrared Detectors for Astronomy
  III}{\nolinebreak\hspace{0.1em}]},  Dorn, D.~A. and Holland, A.~D., eds.,
  {\bf 7021},  54 -- 64, International Society for Optics and Photonics, SPIE
  (2008).

\bibitem{Rauscher_2015}
Rauscher, B.~J., ``Teledyne {H1RG, H2RG, and H4RG} noise generator,'' {\em
  Publications of the Astronomical Society of the Pacific}~{\bf 127}(957),
  1144--1151 (2015).

\bibitem{moore_interpixel_2004}
{Moore}, A.~C., {Ninkov}, Z., and {Forrest}, W.~J., ``{Interpixel capacitance
  in nondestructive focal plane arrays},'' in [{\em Focal Plane Arrays for
  Space Telescopes}{\nolinebreak\hspace{0.1em}]},  {Grycewicz}, T.~J. and
  {McCreight}, C.~R., eds., {\em Society of Photo-Optical Instrumentation
  Engineers (SPIE) Conference Series} {\bf 5167},  204--215 (Jan. 2004).

\bibitem{wang_simultaneously_2021}
Wang, Z., Meng, G., Wang, L., Tian, L., Chen, S., Wu, G., Kong, B., and Cheng,
  Y., ``Simultaneously enhanced dielectric properties and through-plane thermal
  conductivity of epoxy composites with alumina and boron nitride nanosheets,''
  {\em Scientific Reports}~{\bf 11}(1),  2495 (2021).
\newblock Number: 1 Publisher: Nature Publishing Group.

\bibitem{donlon_2018}
Donlon, K., Ninkov, Z., and Baum, S., ``Point-spread function ramifications and
  deconvolution of a signal dependent blur kernel due to interpixel capacitive
  coupling,'' ~{\bf 130}(989),  074503 (2018).

\bibitem{mccullough_2008}
{McCullough}, P., ``{Inter-pixel capacitance: prospects for deconvolution}.''
  Space Telescope WFC Instrument Science Report (2008).

\bibitem{janesick_scientific_2001}
Janesick, J.,  [{\em Photon Transfer: DN $\to
  \lambda$}{\nolinebreak\hspace{0.1em}]}, vol.~PM170, Spie Press Book (2007).

\bibitem{Mortara_1981}
{Mortara}, L. and {Fowler}, A., ``{Evaluations of Charge-Coupled Device / CCD /
  Performance for Astronomical Use},'' in [{\em Society of Photo-Optical
  Instrumentation Engineers (SPIE) Conference
  Series}{\nolinebreak\hspace{0.1em}]},  {Geary}, J.~C. and {Latham}, D.~W.,
  eds., {\em Society of Photo-Optical Instrumentation Engineers (SPIE)
  Conference Series} {\bf 290},  28 (1981).

\bibitem{secroun:hal-01959780}
Secroun, A., Cl{\'e}mens, J.-C., Ealet, A., Gillard, W., Serra, B., Zoubian,
  J., Barbier, R., Ferriol, S., Kubik, B., Rosset, C., Fornari, F., Kohley, R.,
  Conversi, L., Buton, C., and Smadja, G., ``{Euclid flight {H2RG} {IR}
  detectors: per pixel conversion gain from on-ground characterization for the
  Euclid NISP instrument},'' in [{\em {SPIE Astronomical Telescopes +
  Instrumentation 2018}}{\nolinebreak\hspace{0.1em}]},   {\bf 10709},  1070921
  (2018).

\bibitem{mccullough_quantum_2008}
McCullough, P.~R., Regan, M., Bergeron, L., and Lindsay, K., ``Quantum
  {Efficiency} and {Quantum} {Yield} of an {HgCdTe} {Infrared} {Sensor}
  {Array},'' {\em Publications of the Astronomical Society of the Pacific}~{\bf
  120}(869),  759 (2008).

\bibitem{smadja:in2p3-00467907}
Smadja, G., Cerna, C., and Ealet, A., ``{Measurement of the Non-Linearity and
  Interpixel Capacitance of a {H2RG (2Kx2K)} Near-IR Detector},'' {\em {Nuclear
  Instruments and Methods in Physics Research Section A: Accelerators,
  Spectrometers, Detectors and Associated Equipment}}~{\bf 610},  615--621
  (2009).

\bibitem{hirata_2019}
Hirata, C.~M. and Choi, A., ``Brighter-fatter effect in near-infrared
  detectors. i. theory of flat autocorrelations,'' {\em Publications of the
  Astronomical Society of the Pacific}~{\bf 132}(1007),  014501 (2019).

\bibitem{kubik:in2p3-01388431}
Kubik, B., Barbier, R., Chabanat, E., Chapon, A., Clemens, J.-C., Ealet, A.,
  Ferriol, S., Gillard, W., Secroun, A., Serra, B., Smadja, G., and Tilquin,
  A., ``{A New Signal Estimator from the NIR Detectors of the Euclid
  Mission},'' {\em {Publications of the Astronomical Society of the
  Pacific}}~{\bf 128}(968),  104504 (2016).

\end{thebibliography}
\bibliographystyle{spiebib} 

\end{document}